\documentclass[a4paper,12pt,tightenlines,aps,showpacs]{revtex4}

\usepackage{amsfonts}
\usepackage{amsmath}
\usepackage{amssymb}
\usepackage[american]{babel}
\usepackage{graphicx}
\usepackage{subfigure}
\usepackage{multirow}

\newcommand{\mat}{\left( \begin{array}{cc}}
\newcommand{\rix}{\end{array} \right)}

\newcommand{\Z}{\mathbb{Z}}

\newcommand{\eps}{\epsilon}

\newcommand{\e}{{\rm e}}

\newcommand{\norm}[1]{\left\| #1 \right\|}

\newtheorem{definition}{Definition}[section]

\newcommand{\ket}[1]{|#1\rangle}
\newcommand{\bra}[1]{\langle#1|}

\newcommand{\scalar}[2]{\langle#1|#2\rangle}

\begin{document}

\title{Mixing Times in Quantum Walks on Two-Dimensional Grids}
\author{F.L.~Marquezino$^1$, R.~Portugal$^1$ and G. Abal$^2$}
%\email{franklin@lncc.br} %\email{portugal@lncc.br} %\email{abal@fing.edu.uy}
\affiliation{$^1$Laborat\'{o}rio Nacional de Computa\c{c}\~{a}o
  Cient\'{\i}fica - LNCC\\Avenida  Get\'{u}lio Vargas 333, Petr\'{o}polis,
  RJ, 25651-075, Brazil}
\affiliation{$^2$Instituto de F\'{\i}sica, Universidad de la Rep\'ublica\\
   Casilla de Correo 30, C\'odigo Postal 11300, Montevideo, Uruguay}

\begin{abstract}
Mixing properties of discrete-time quantum walks on
  two-dimensional grids with torus-like boundary conditions
  are analyzed, focusing on their connection to the complexity
  of the corresponding abstract search algorithm.  In particular,
  an exact expression for the stationary distribution of the coherent walk over odd-sided
  lattices is obtained after solving the eigenproblem for the evolution
  operator for this particular graph.  The limiting distribution and
  mixing time of a quantum walk with a coin operator modified as in the abstract search
  algorithm are obtained numerically. On the basis of these results, the relation between the mixing
  time of the modified walk and the running time of the corresponding abstract search algorithm is discussed.
\end{abstract}
\pacs{03.67.Lx, 05.40.Fb, 03.65.Yz}

\maketitle

\section{Introduction}

The concept of quantum walk is analogous to that of classical random
walks, with the walker replaced by a quantum particle and coherent
superpositions playing a key role~\cite{Kempe03}. There are
discrete-time and a continuous-time versions of the quantum walk---the
former defined by Aharonov \textit{et al.}~\cite{Davidovich} and the
latter, by Farhi and Gutmann~\cite{FG98}. Similarly to random walks,
which have been used as a basis for classical algorithms that
outperform their deterministic counterparts~\cite{Motwani1995},
quantum walks have also been used as a basis for quantum algorithms
that outperform their classical
correspondents~\cite{Shenvi,Amb03,CG04,Farhi07}.

Aharonov \textit{et al}~\cite{Aharonov} have presented important
results on the theory of quantum walks on graphs, several of them
concerning mixing-time properties. In particular, they have analyzed odd-sided $N$-cycles
and showed that the quantum walk converges to a stationary distribution in time $O(n \log n)$,
almost quadratically faster than the classical walk. They have also obtained
bounds on mixing-times for general graphs. Mixing times on hypercubes
were properly addressed by Marquezino \textit{et al}~\cite{marquezino}, who presented an analytical
expression for the stationary distribution of a coherent discrete-time
quantum walk on the hypercube, a Cayley graph with important algorithmic applications~\cite{Shenvi}. The stationary distribution is the first step in the calculation of the mixing time. An important graph is missing in the mixing-time
picture: the two-dimensional grid.  This grid, with torus-like boundary conditions, is the natural extension of the $N$-cycle to two dimensions. The quantum search in this graph is one of the first examples of abstract search algorithms~\cite{AKR,Tulsi2008}, a general framework for developing and analyzing quantum walk searches on graphs.

In this paper, we consider the mixing time of a discrete-time quantum walk on torus-like 2D-grids, and analyze the
relation between these properties and the complexity of the
corresponding abstract search. First, the eigenvector problem of the evolution
operator for the quantum walk on the two-dimensional torus is solved. This is an essential result for the analysis of many mathematical properties of the quantum walk. Then, the limiting probability distribution of the coherent walk in odd-sided lattices is derived, for an initial condition localized at the origin in a uniform superposition of coin states. In this case, the limiting probability distribution is found to have a maximum at the origin. The mixing time has been numerically calculated and we show that the walk mixes in time $O(\sqrt{N \log N})$, where $N$ is the total number of vertices.

A quantum walk in which the coin is modified according to the prescription of the abstract search
algorithm has also been considered. Its limiting distribution and its mixing time have been numerically obtained. These results imply a relation between the mixing time of the modified walk and the complexity of the
corresponding abstract search algorithm. In previous works, the complexity of the quantum search algorithm was considered to be related only to the hitting time, while the mixing time was considered to be related to data sampling~\cite{Richter}. We show how previous knowledge of the limiting distribution of the quantum walk may be used to estimate the running time of an abstract search algorithm.

The paper is organized as follows.  In Section~\ref{sec:coherent}, we
consider the coherent quantum walk on the two-dimensional finite grid
with torus-like boundary conditions and solve the eigenvector problem
of its evolution operator.  In Section~\ref{ssec:limit-dist}, the limiting distribution of the coherent walk for the case of odd-sided lattices is derived.  In Section~\ref{ssec:coh-mix-time}, the results of numerical simulations used to estimate the mixing time for the coherent evolution are presented.  In Section~\ref{ssec:MTASA}, the relation between the mixing time and the running time of the abstract search algorithm on the two-dimensional grid is discussed.  In
Section~\ref{sec:conclusion}, the main results are summarized and
our conclusions are presented.

\section{Coherent walk on a two-dimensional grid}
\label{sec:coherent}

A coined quantum walk in a $\sqrt N\times \sqrt N$ grid with
periodical boundary conditions has a Hilbert space ${\cal H}_C\otimes
{\cal H}_P$, where ${\cal H}_C$ is the $4$-dimensional coin subspace
and ${\cal H}_P$ the $N$-dimensional position subspace. A basis for
${\cal H}_C$ is the set $\{\ket{d,s}\}$ for $0\leq d,s\leq 1$ and
${\cal H}_P$ is spanned by the set $\{\ket{x,y}\}$ with $0 \leq
x,y\leq \sqrt N$, where we assume integer $\sqrt{N}$. A generic state of the quantum walk is
\begin{equation}
  \ket{\Psi(t)}=\sum_{d,s=0}^{1}\sum_{x,y=0}^{\sqrt N-1}\psi_{d,s;x,y}(t)\ket{d,s}\ket{x,y}. \label{eq:estgeral1d}
\end{equation}
The evolution operator for one step of the walk is
\begin{equation}\label{evol}
  U=S\cdot(C\otimes I),
\end{equation}
where $I$ is the identity in ${\cal H}_P$, $S$ is the shift operator
\begin{equation}\label{shift-coh}
  S=\sum_{d,s=0}^{1}\sum_{x,y=0}^{\sqrt N-1}\ket{d,s\oplus 1}\bra{d,s}\otimes \ket{x+(-1)^s\delta_{d 0},y+(-1)^s\delta_{d 1}}\bra{x,y}
\end{equation}
and $C$ is a unitary coin operation in ${\cal H}_C$. Notice that the binary sum $\oplus$ in $S$ inverts the direction, as required by the abstract search algorithm~\cite{AKR}. We shall conisder the Grover coin, $C=G$, given by
\begin{equation}\label{GroverCoin}
  G = {2 \ket{u}\bra{u}-I},
\end{equation}
where $\ket{u}=\frac{1}{2}\sum_{d,s=0}^1\ket{d,s}$ is the uniform superposition in ${\cal H}_C$.

\subsection{Eigenproblem for $U$}
\label{ssec:eigenp-U}

The analysis of the problem is simplified in Fourier space. Since
the two-dimensional grid with periodic boundary conditions is a Cayley
graph of $\Z_{\sqrt N}^2$, we use the Fourier transform on this group,
which has a basis spanned by the $N$ kets
\begin{equation}
 \ket{k_x, k_y}=\frac{1}{\sqrt N}\sum_{x,y=0}^{\sqrt{N}-1} \omega^{x k_x + y k_y} \ket{x,y},
\end{equation}
where $\omega =\e^\frac{2\pi i}{\sqrt N}$.  The components of the
evolution operator in the Fourier space are
\begin{equation}
  \bra{d,s,k_x^\prime,k_y^\prime} U \ket{d^\prime,s^\prime,k_x,k_y}= \omega ^{(-1)^s(\delta_{d 0} k_x + \delta_{d 1} k_y)} G_{d,s\oplus 1;\,d^\prime,s^\prime} \, \delta_{k_x,k_x^\prime}\delta_{k_y,k_y^\prime}.
\end{equation}
For each $k_x,k_y$, we define a reduced evolution operator in the coin subspace given by
\begin{equation}
  \widetilde{G}_{d,s;\,d^\prime,s^\prime}=\omega ^{(-1)^s(\delta_{d 0} k_x + \delta_{d 1} k_y)} G_{d,s\oplus 1;\,d^\prime,s^\prime},
\end{equation}
which is a $4\times 4$ matrix that can be diagonalized. The eigenvectors of $U$ are tensor products of the eigenvectors of $\widetilde{G}$ and $\ket{k_x,k_y}$.

Let us now describe the eigenspectrum of $\widetilde{G}$. If $k_x=0$ and
$k_y=0$, the  eigenvalue 1 is three-fold degenerate and the eigenvectors are
$\frac{1}{\sqrt 2}(1,-1,0,0)^T$, $\frac{1}{\sqrt 2}(0,0,1,-1)^T$ and
$\ket{u} \equiv \frac{1}{2}(1,1,1,1)^T$, where $(\ldots)^T$ is a column vector. The eigenvector with
eigenvalue $-1$ is $\frac{1}{2}(1,1,-1,-1)^T$. If $k_x\neq 0$ or $k_y\neq 0$, the eigenvalues are $\pm 1$ and $e^{\pm i\theta}$ with $\theta$ defined by \begin{equation}
  \label{cos theta}
  \cos \theta=\frac{1}{2} \left[ \cos\left({\frac {2 \pi{k_x}}{\sqrt {N}}}\right) +\cos\left({\frac {2 \pi{k_y}}{\sqrt{N}}}\right)\right].
\end{equation}
In this case, the eigenvectors of $\widetilde{G}$ with eigenvalue $+1$ are
\begin{equation}
  \label{pq_nu_plus1} %%%%2\sqrt{2}\sqrt{1-\cos \theta}
  \ket{\nu^{+1}_{k_x,k_y}}=\frac{1}{4\sin(\theta/2)}
  \begin{bmatrix}
    \omega^{k_x}\left( \omega^{k_y}-1 \right) \\
    1-\omega^{k_y}\\
    \omega^{k_y} \left( 1-\omega^{k_x} \right) \\
    \omega^{k_x}-1
  \end{bmatrix}
\end{equation}
and those with eigenvalue $-1$ are
\begin{equation}
  \label{pq_nu_minus1}
  \ket{\nu^{-1}_{k_x,k_y}} = \frac{1}{4\cos (\theta/2)}
  \begin{bmatrix}
    -\omega^{k_x}\left( 1+\omega^{k_y} \right) \\
    -\left( 1+\omega^{k_y} \right) \\
    \omega^{k_y} \left( 1+\omega^{k_x} \right) \\
    1+\omega^{k_x}
  \end{bmatrix}.
\end{equation}
Finally, the eigenvectors with eigenvalues $\e^{i\theta}$ are
\begin{equation}\label{pq_nu_theta}
  \ket{\nu^{+\theta}_{k_x,k_y}}=
  \frac{i}{2\sqrt 2 \sin \theta}
  \begin{bmatrix}
    \e^{-i\theta}-\omega^{k_x}\\
    \e^{-i\theta}-\omega^{-k_x}\\
    \e^{-i\theta}-\omega^{k_y}\\
    \e^{-i\theta}-\omega^{-k_y}
  \end{bmatrix}.
\end{equation}
and the eigenvectors with eigenvalue $\e^{-i\theta}$ are obtained by the replacing $\theta\rightarrow -\theta$ in Eq.~(\ref{pq_nu_theta}). Note that the eigenvectors are normalized, $\ket{\nu^{\pm\theta}_{k_x,k_y}}$, form an orthonormal basis for the reduced space and have a real constant component on the uniform state,  $\scalar{\nu^{\pm\theta}_{k_x,k_y}}{u}=1/\sqrt 2$.

We take the state
\begin{equation}
  \label{IC}
  \ket{\Psi(0)}=\ket{u}\ket{x=0,y=0}
\end{equation}
as the initial condition, i.e., the walker starts localized at the
point $(0,0)$ and uniformly distributed in the coin subspace. In the
eigenbasis, this initial condition is given by
\begin{equation}
  \label{eq:ini_cond}
  \ket{\Psi(0)} = \frac{1}{\sqrt N}\,\ket{u}\ket{k_x=0,k_y=0} +
  \frac{1}{\sqrt{2N}}\sum_{
    \begin{subarray}{c}
      {k_x,k_y=0}\\
      (k_x, k_y)\neq (0,0)
    \end{subarray}}^{\sqrt N -1}
  \left(\ket{\nu^{+\theta}_{k_x,k_y}}+
  \ket{\nu^{-\theta}_{k_x,k_y}}\right)\ket{k_x,k_y}.
\end{equation}
Applying $U^t$ on $\ket{\Psi(0)}$ we obtain the state of the quantum walk after $t$ steps,
\begin{multline}
  \ket{\Psi(t)} = \frac{1}{\sqrt N}\,\ket{u}\ket{0,0} +
  \frac{1}{\sqrt{2N}}\sum_{
    \begin{subarray}{c}
      {k_x,k_y=0}\\
      (k_x, k_y)\neq (0,0)
    \end{subarray}}^{\sqrt N -1} \left(\e^{i\theta t} \,
  \ket{\nu^{+\theta}_{k_x,k_y}} + \e^{-i\theta
    t} \, \ket{\nu^{-\theta}_{k_x,k_y}}\right)\ket{k_x,k_y}.
  \label{eq:state-time-t}
\end{multline}
Having solved the eigenvector problem for the evolution operator, we proceed to the
calculation of the limiting distribution for the quantum walk.

\subsection{Limiting Distribution}
\label{ssec:limit-dist}

Let $P(x,y, t)$ be the probability to find the walker at a vertex
$(x,y)$ of the grid at time $t$. As mentioned in the introduction,
this probability depends on the initial condition and, as is typical
of unitary evolutions, it does not converge to a stationary
distribution. However, the time-averaged distribution $\protect{\bar P
  (x,y,T)\equiv\frac{1}{T}\sum_{t=0}^{T-1} P(x,y, t)}$ always converges
as
% $T\to\infty$
$T$ goes to infinity~\cite{Aharonov}. The limiting or stationary
distribution is then defined in terms of the average distribution,
\begin{equation}
  \pi(x,y)\equiv\lim_{T\to\infty} \bar P(x,y,T).
  \label{eq:Pbar}
\end{equation}%

Using Theorem~3.4 from Ref.~\cite{Aharonov}, the coefficients of the
initial condition expressed in the eigenbasis of $U$, Eq.~(\ref{eq:ini_cond}),  and the fact that
$\scalar{{\nu_{k_x^\prime,k_y^\prime}^{+\theta^\prime}}}{\nu_{k_x,k_y}^{+\theta}}=
\scalar{{\nu_{k_x^\prime,k_y^\prime}^{-\theta^\prime}}}{\nu_{k_x,k_y}^{-\theta}}$,
we obtain a simple expression for the limiting distribution,

\begin{equation}\label{eq:pi}
  \pi({x,y})=\frac{1}{N^2} + \frac{1}{N^2}\sum_{
    \begin{subarray}{c}
      {(k_x,k_y)\neq (0,0)}\\
      {(k_x^\prime, k_y^\prime)\neq (0,0)}\\
      {\theta(k_x,k_y)=\theta(k_x^\prime, k_y^\prime)}
    \end{subarray}}
  \scalar{\nu_{k_x^\prime,k_y^\prime}^{\theta}}{\nu_{k_x,k_y}^{\theta}}\,\,
  \omega^{x(k_x-k_x^\prime)+y(k_y-k_y^\prime)}.
\end{equation}
When $\theta(k_x^\prime,k_y^\prime)=\theta(k_x,k_y)$ we have
\begin{equation}
  \scalar{ {\nu_{k_x^\prime,k_y^\prime}^{\theta}} }{ \nu_{k_x,k_y}^{\theta} } = \frac{1 - 2\cos^2\left[ \theta(k_x,k_y)\right]+\cos\left[ \theta(k_x-k_x^\prime,k_y-k_y^\prime)\right]}{2\sin^2\left[\theta(k_x,k_y)\right]}.
\end{equation}

Analyzing all cases such that
$\theta(k_x^\prime,k_y^\prime)=\theta(k_x,k_y)$, the expression for
the limiting distribution, for odd $\sqrt N$, has the explicit form
\begin{eqnarray}
  \pi(x,y) = \frac{1}{N} &+&
  \frac{2}{N^2}\sum_{k_x=1}^{\sqrt{N}-1} \frac{1}{3+\cos\tilde{k}_x} \nonumber\\
 &\times& \Biggl\{ \left[ (x-y)\cos {\tilde{k}_x} + \omega^{k_x(x+y)} \right]
  \left({1+\cos\tilde{k}_x}\right) + 2\left( \omega^{2k_x x} + \omega^{2k_x y} \right)\Biggr\} + \nonumber\\
  &+&\frac{1}{2N^2}\sum_{\stackrel{k_x,k_y=1}{k_y\not\in\{k_x,\sqrt{N}-k_x\}}}^{\sqrt{N}-1} \frac{1}{\sin^2\theta}
  \Biggl\{
   \left[ \cos {(\tilde{k}_x-\tilde{k}_y)} - \cos 2\theta  \right] \omega^{(k_x-k_y)x+(k_y-k_x)y} \nonumber\\
  &&\quad +\left[ \cos^2\tilde{k}_x - \cos 2\theta \right] \omega^{2k_x x}
   \left[ \cos^2\tilde{k}_y - \cos 2\theta \right]\omega^{2k_y y} +\label{eq:limiting}\\
  &&\quad\quad +\left[ \cos\left(\theta(k_x-k_y,k_x+k_y)\right) - \cos 2\theta \right]\omega^{(k_x-k_y)x+(k_x+k_y)y} \nonumber\\
  &&\quad\quad\quad +\frac{1}{2}\left[ \sin^2\tilde{k}_x + \sin^2\tilde{k}_y\right]\omega^{(k_x+k_y)x+(k_y-k_x)y} \nonumber\\
  &&\quad\quad\quad\quad + \frac{1}{2}\left[ \cos\tilde{k}_x - \cos\tilde{k}_y \right]^2 \omega^{2k_x x+2k_y y}\nonumber\\
  &&\quad\quad\quad\quad\quad +\left[ \cos {(\tilde{k}_x+\tilde{k}_y)} - \cos 2\theta \right] \omega^{(k_x+k_y)x+(k_x+k_y)y}\Biggr\},\nonumber
\end{eqnarray}
where $\tilde{k}_x=2\pi k_x/\sqrt{N}$ and $\tilde{k}_y=2\pi k_y/\sqrt{N}$.

\begin{figure}[h]
  \centering
  \includegraphics[height=0.5\textwidth,angle=270]{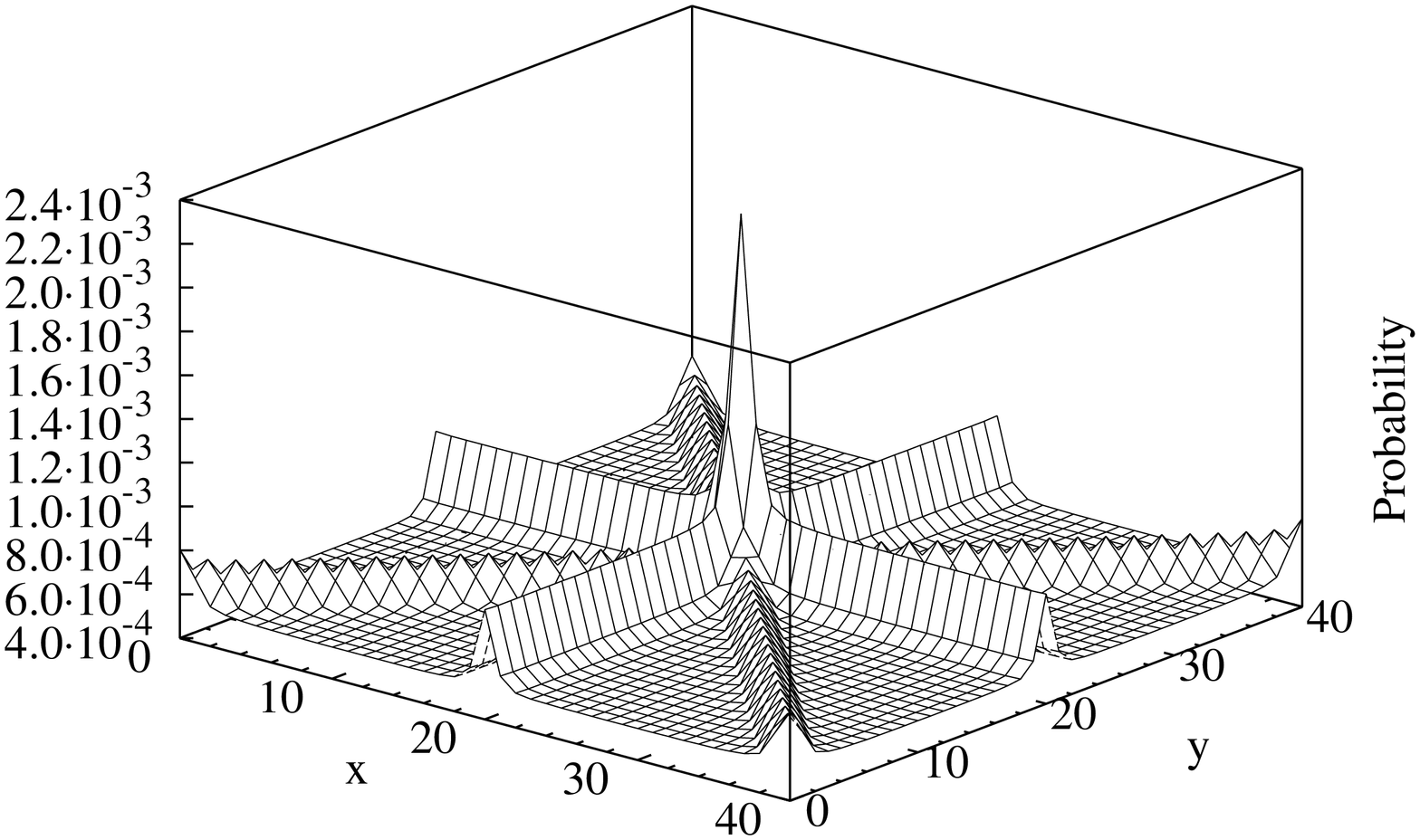}~
  \includegraphics[height=0.45\textwidth,angle=270]{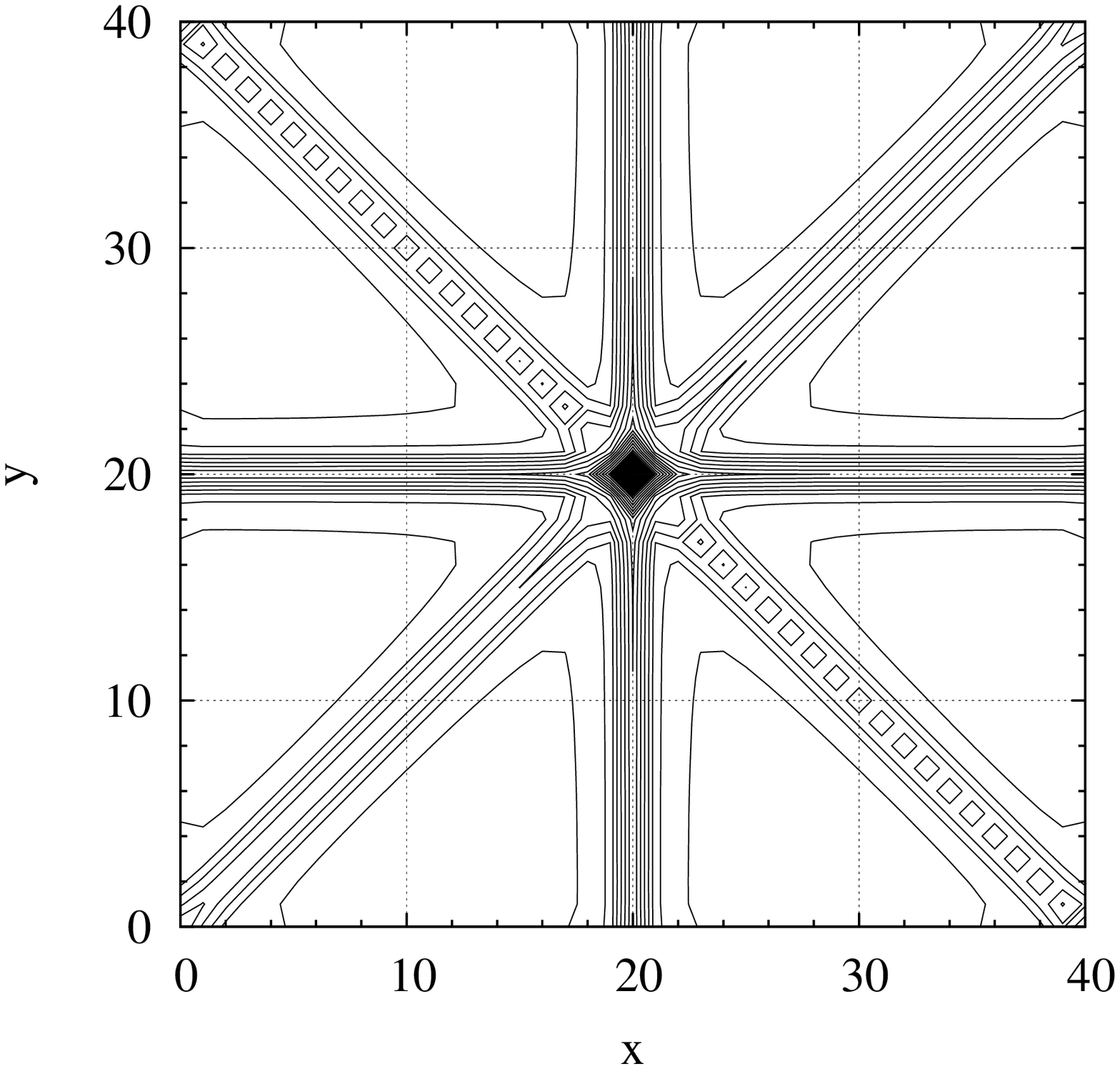}
  \caption{Left panel: Limiting distribution for quantum walk in
    two-dimensional grid with $\sqrt{N}=41$, obtained from
    Eq.~\eqref{eq:limiting} with the initial
    condition~\eqref{IC}. Right panel: contour plot for the same
    distribution.}
  \label{fig:n101}
\end{figure}

For some values of $(x,y)$, it is possible to simplify
Eq.~\eqref{eq:limiting} and achieve simple results. An interesting
point to be considered is the initial site, which was
assumed to be $(x_0,y_0)=(0,0)$, without loss
of generality. It is straightforward to show that
\begin{equation}
  \pi(0,0) = \frac{4 N-{8}{\sqrt N}+{5}}{N^2}
\end{equation}
and, for $N\gg 1$, we obtain $\pi(0,0) \approx 4/N$, which is also the maximum of the limiting distribution. This distribution is shown  in Fig.~\ref{fig:n101} for a grid of dimensions $41\times 41$, with the initial site
shifted to $(x_0,y_0)=(20,20)$ for better visualization.

It is well known that the behavior of quantum walks on even lattices
may be different than that observed for odd
lattices~\cite{Aharonov,KT02}.  The analysis of the walk over odd
lattices is sufficient for the main objective of this paper, namely to
study the relation between mixing time and the complexity of
quantum-walk based search algorithms.  Our numerical simulations show
that the liming distribution for even lattices present two peaks,
instead of the single peak on the initial site as observed for the odd
lattice.
Now that we have the limiting distribution for the quantum walk on the
two-dimensional grid, we consider its mixing time and its relation to the abstract search problem.

\section{Mixing times and abstract search}

In this subsection we consider the mixing time for a coherent
evolution. The rate at which the average probability distribution of a
quantum walk approaches its asymptotic distribution is captured by the following
definition \cite{Aharonov}.

\begin{definition}
  The average mixing time $M_\epsilon$ of a quantum Markov chain to a
  reference distribution $\pi$ is
$$M_\epsilon=\min\{T\,|\,\forall t\ge T,\norm{\bar{P}_t - \pi} \leq \epsilon \},$$ where
$\norm{A - B}\equiv \sum_x |A(x) - B(x)|$ is the total variation
distance between the two distributions.
\end{definition}
An alternative definition captures the first instant in which the walk
is $\epsilon$-close to the reference distribution $\pi$,
\begin{definition}
  The instantaneous mixing time $I_\epsilon$ of a quantum Markov chain
  is
$$
I_\epsilon=\min\{t\,|\,\norm{P_t - \pi} \leq \epsilon\}.
$$
\end{definition}
Both mixing times depend on the initial condition of the quantum walk.

For a quantum walk in a generic graph with arbitrary initial
condition, an upper bound for the total variation distance to the
asymptotic distribution $\pi(x,y)$ was derived by Aharonov \textit{et
  al.}~\cite{Aharonov},
\begin{equation}\label{eq:theo6.1}
  \|\bar P(x,y,T) - \pi(x,y)\|\leq \frac{\pi}{T\Delta}\left[\log\left(\frac{Nd}{2}+1\right)\right],
\end{equation}
where $N$ is the number of vertices, $d$ is the degree of each vertex
and $\Delta$ is the minimum separation between distinct eigenvalues of
$U$.

\subsection{Mixing time in the two-dimensional grid}
\label{ssec:coh-mix-time}

\begin{figure}[h]
  \centering
  \includegraphics[width=0.45\textwidth]{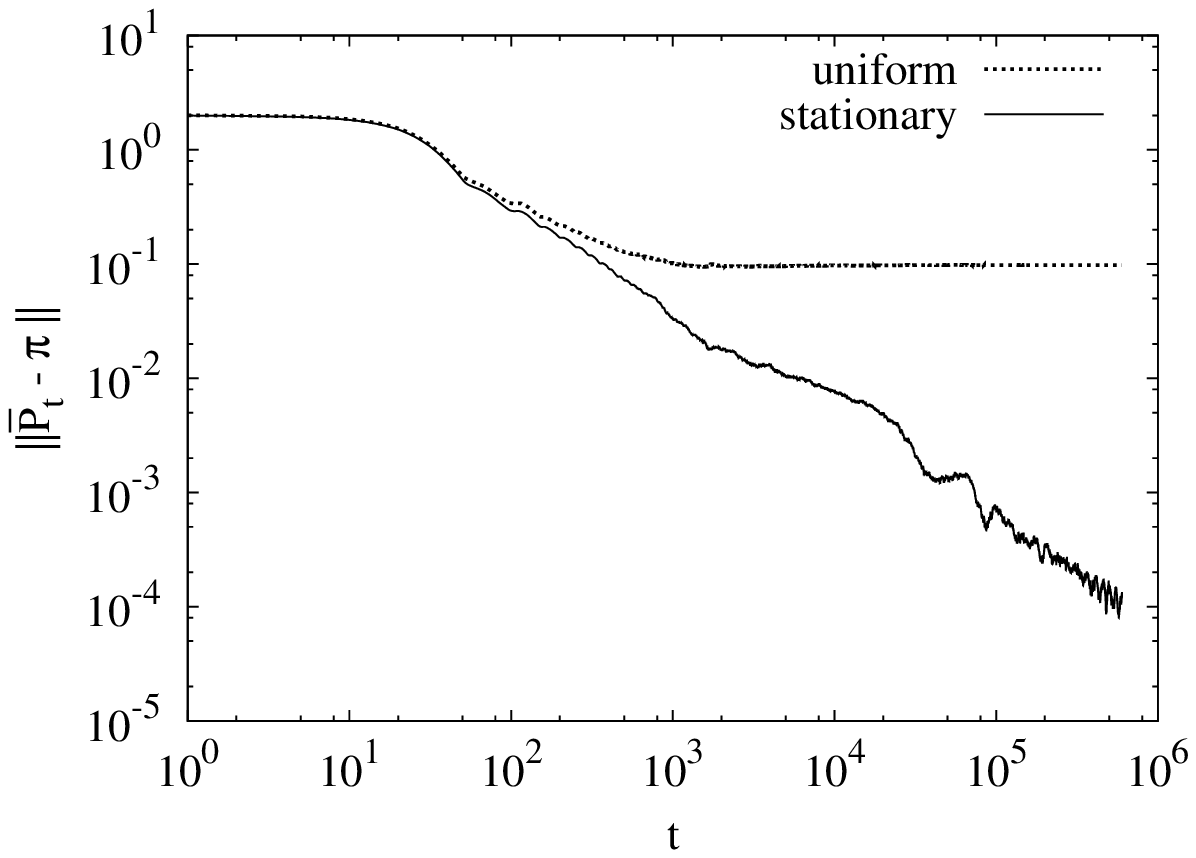}~
  \includegraphics[width=0.45\textwidth]{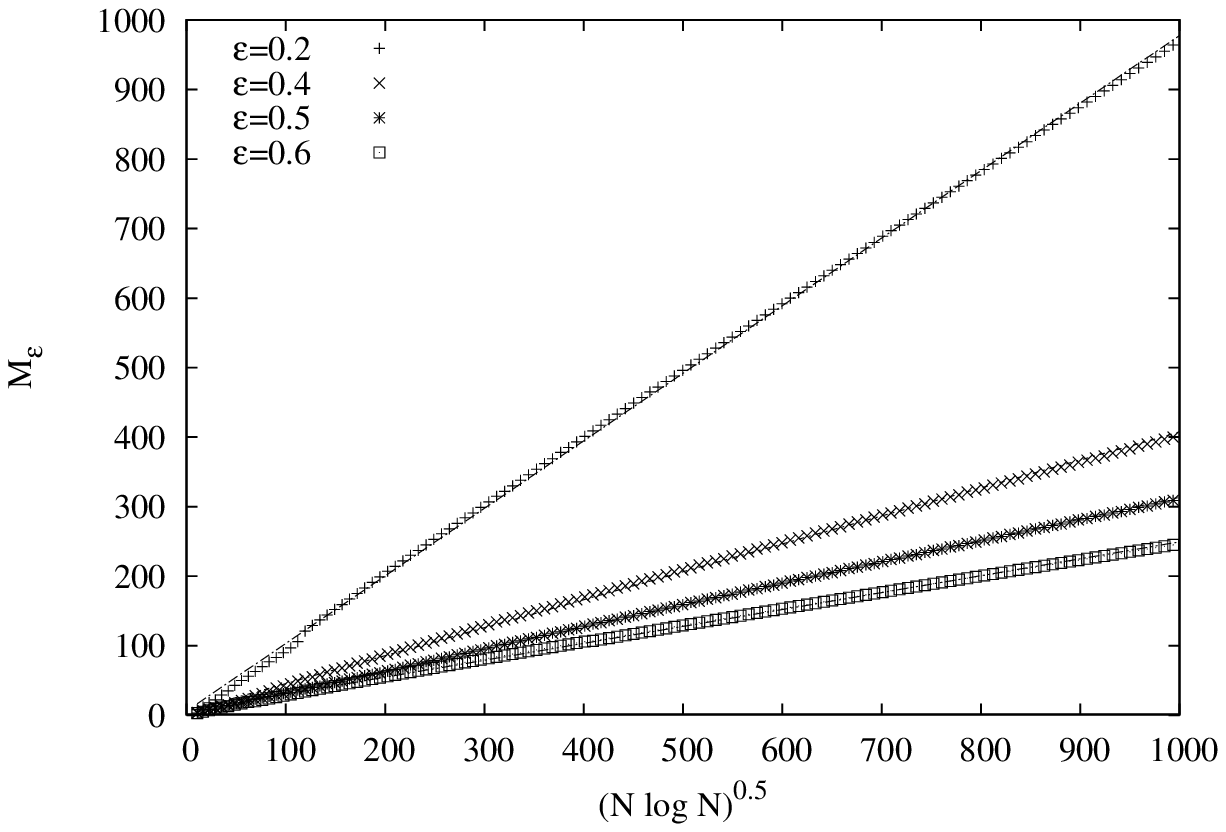}
  \caption{Left panel: total variation distance to both the uniform
    and the stationary distributions of the quantum walk on the
    two-dimensional grid with flip-flop shift as a function of the
    time step. Right panel: mixing time to the stationary distribution
    as a function of the input size.}
  \label{fig:preakr}
\end{figure}

Let us now consider the particular case of a two-dimensional cartesian grid, for which $d=4$.
The value of $\Delta$ is the minimum value of
$\left|\e^{i\theta(k_x,k_y)}-\e^{i\theta(k_x^\prime,k_y^\prime)}\right|$
for $k_x,k_y$ and $k_x^\prime,k_y^\prime$ in the range
$\big[0,\sqrt{N}-1\big]$. For large $N$ and small values of $k_x$,
$k_y$, $k_x^\prime$ and $k_y^\prime$, we obtain
\begin{equation}\label{eq:asymptDelta}
  \Delta\approx \frac{\sqrt 2 \pi}{\sqrt N}\left|\sqrt{k_x^2+k_y^2}-\sqrt{k_x^{\prime \, 2}+k_y^{\prime\,2}}\,\right|.
\end{equation}
The minimum value of $\Delta$ is obtained taking values of $k_x$,
$k_y$, $k_x^\prime$ and $k_y^\prime$ such that
$\theta(k_x,k_y)\neq\theta(k_x^\prime,k_y^\prime)$ and
\begin{equation*}
  \sqrt{k_x^2+k_y^2}\approx\sqrt{k_x^{\prime \, 2}+k_y^{\prime\,2}}.
\end{equation*}
We can make a good guess by exploring the structure of the above
equation. With those guesses we can try to find an upper bound for the average mixing time,
$M_\epsilon$. It is easy to obtain the value~$O(1/\sqrt N)$
for~$\Delta$.  Therefore, a good guess for $M_\epsilon$ is
$O\left(\frac{\sqrt N\log N}{\epsilon}\right)$.  By performing a
numerical analysis, we have succeeded in obtaining better bounds regarding the dependence on $N$.
Plotting $M_\epsilon$ against $\sqrt{N\log N}$ for several values of
$\epsilon$ we obtain straight lines as shown in the right
panel of Fig.~\ref{fig:preakr}. These results show that $M_\epsilon$ is proportional to
$1/\epsilon^c$, where $c$ is approximately 1. Therefore, the numerical data strongly suggests
that
\begin{equation}\label{eq:Mepsilon2}
  M_\epsilon=\Theta\left(\frac{\sqrt{N\log N}}{\epsilon^c}\right).
\end{equation}

In the left panel of Fig.~\ref{fig:preakr}, we have the total
variation distance to both the uniform and the stationary
distribution. For long times, the variation distance to the stationary
distribution decays approximately as~$\sim 1/t$ while the
corresponding distance to the uniform distribution remains essentially
constant.

The classical random walk on a two-dimensional grid increases with the
size of the lattice as $\Theta(N)$---see, for instance, chapters 3 and 5 of
Aldous and Fill~\cite{Aldous}. Comparing this result with
Eq.~\eqref{eq:Mepsilon2}, we observe that the quantum walk mixes
almost quadratically faster than its classical counterpart on the same
lattice. The faster mixing rates observed in the quantum case is one
of the main advantages of using quantum algorithms over their classical
equivalents.

Until now, we have analyzed the standard quantum walk on the torus, without
direct algorithmic applications. In the next section, we investigate
the behavior of a quantum walk with a modified evolution operator which makes
it useful to to mark a searched vertex of the grid.

\subsection{Connection with abstract search algorithms}
\label{ssec:MTASA}

In this section we discuss the relation between the mixing time and
the running time of the abstract search algorithm on the
two-dimensional grid.  The running time of search algorithms is
usually associated with the notion of hitting time~\cite{Szegedy},
which is defined as the first time a given vertex is reached.  On the
other hand, the mixing time is usually associated with data
sampling~\cite{Richter}.  We argue that those measures are related and
both may be used to estimate the running time of abstract search
algorithms.

The abstract search algorithm is a search framework introduced by
AKR~\cite{AKR} based on a modified quantum walk. The standard quantum walk
is driven by the evolution operator given by Eq.~(\ref{evol}). The
modified operator is $U^\prime=S\cdot C^\prime$, where $C^\prime$ is
given by
\begin{equation}\label{Cprime}
  C^\prime= -I\otimes \ket{x_0,y_0}\bra{x_0,y_0} + G\otimes \big(I- \ket{x_0,y_0}\bra{x_0,y_0}\big).
\end{equation}
The modified operator $C^\prime$ applies the coin $-I$ if the vertex is the target $\ket{x_0,y_0}$, otherwise it applies Grover's coin operation $G$. The effect of $C^\prime$ is to mark the searched vertex with a
relative phase. With this new evolution operator, AKR have shown that
a quantum walker departing from the uniform distribution will be at the
marked vertex after $O(\sqrt{ N \log N})$ steps with probability
$O(1/\log N)$. The time complexity of AKR's algorithm is $O(\sqrt{ N }\,{\log N})$,
after using the method of amplitude amplification~\cite{KLM07}.
Tulsi~\cite{Tulsi2008} has improved this result
and have shown a method to reach the marked vertex with probability $O(1)$.
The time complexity of Tulsi's algorithm is $O(\sqrt{ N \log N})$.

The effect of the modified evolution operator is to increase the
probability of finding the walker in the marked vertex at some
specific steps. The best point to stop is the one at which the
probability is maximum. AKR described a method to find the running
time without calculating directly the point of maximum and without
using the notion of hitting time. On the other hand,
Szegedy~\cite{Szegedy} has described a general method to obtain the
hitting time. Szegedy's method, however, requires a change on the
evolution operator, which does not have the same time complexity of
AKR's method for searching one vertex on the two-dimensional grid.

% Graph for the limiting distribution

\begin{figure}[h]
  \centering
  \includegraphics[height=0.45\textwidth,angle=270]{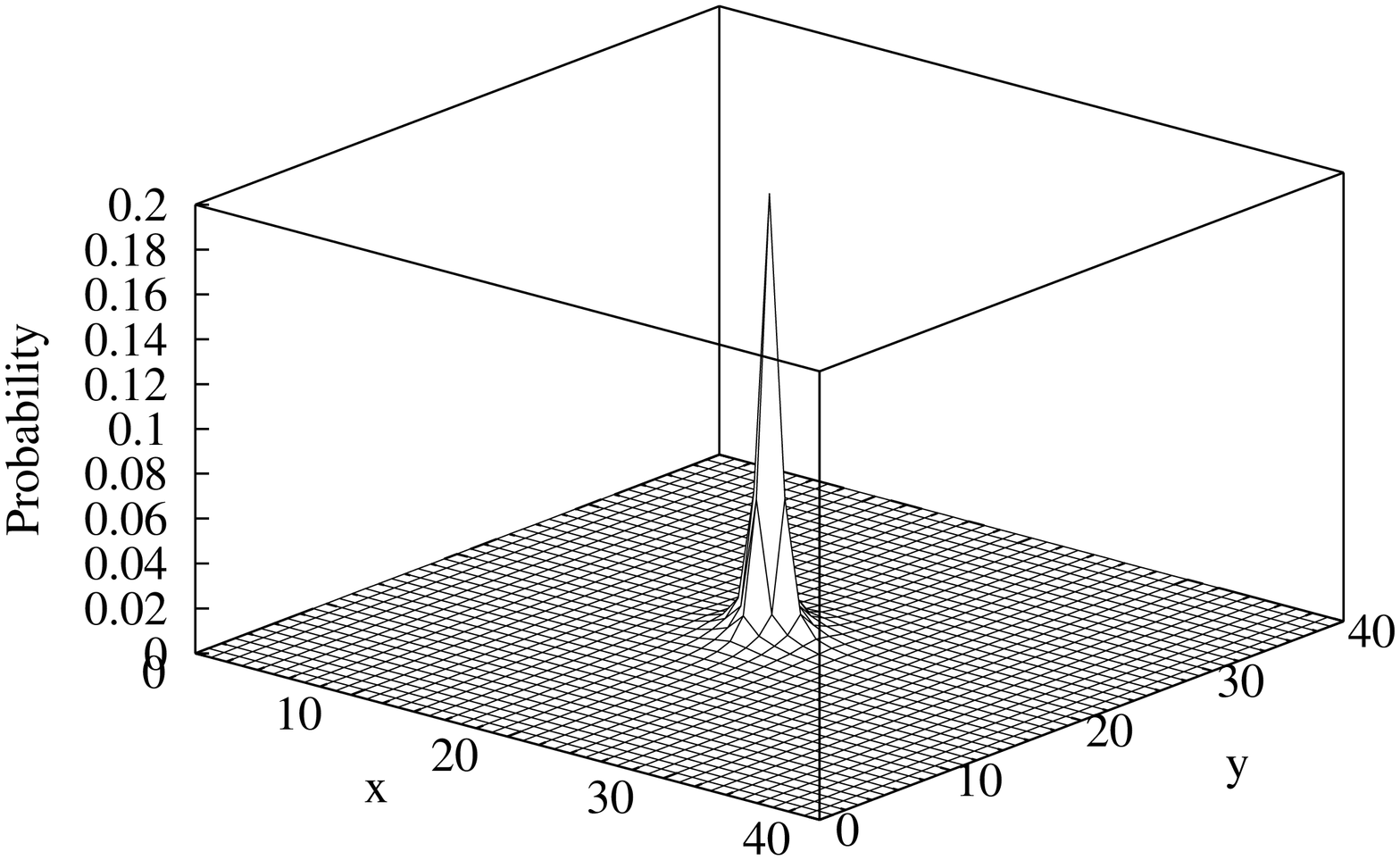}~
  \includegraphics[height=0.45\textwidth,angle=270]{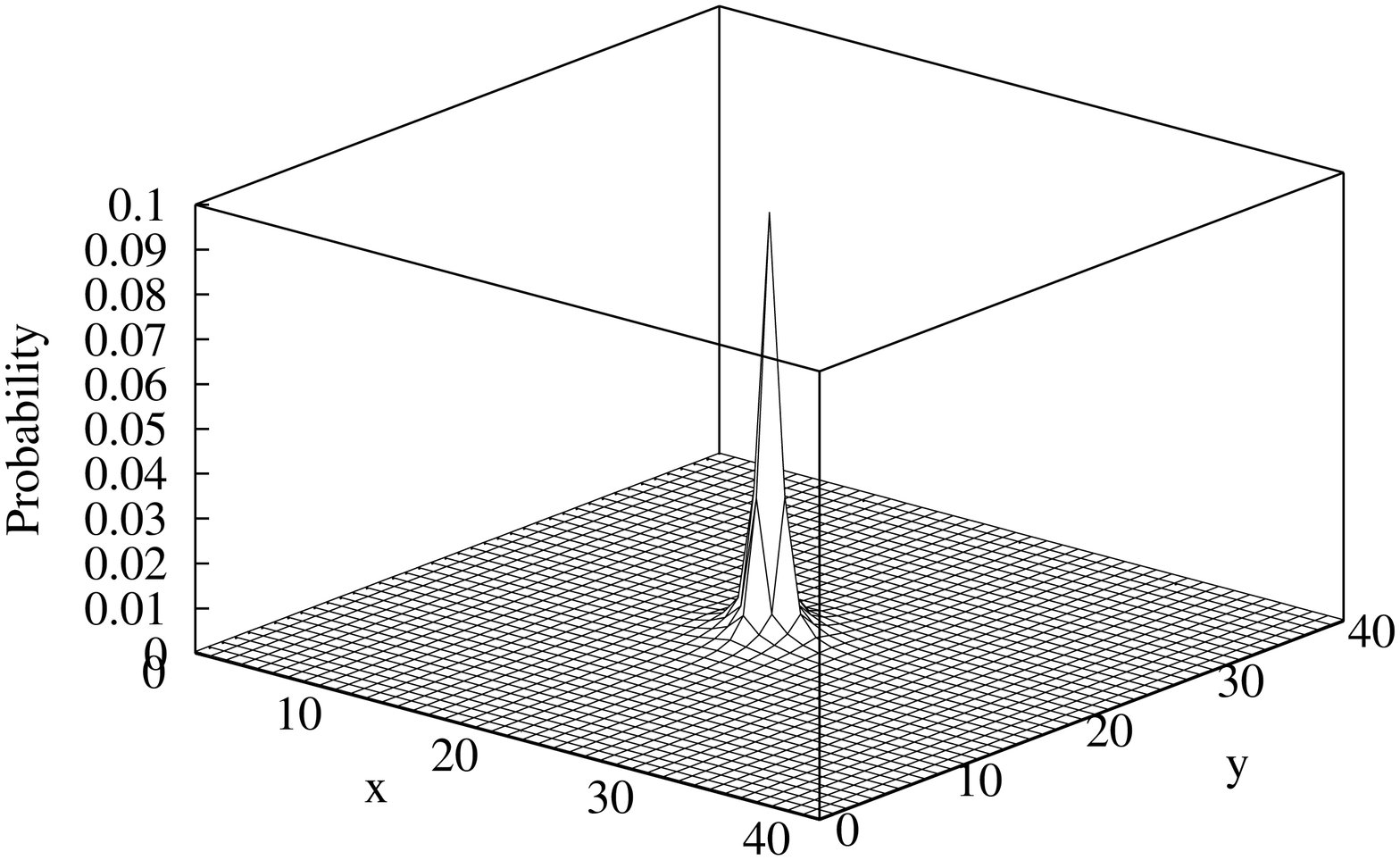}
  \caption{Quantum walk in two-dimensional grid with $\sqrt{N}=41$ and a
    modified coin used to search for a marked node. Left panel:
    Probability distribution after~$t=80$ steps, corresponding to the
    instant of maximum probability at the marked node. Right panel:
    stationary distribution approximated with $T=10^4$ simulation
    steps.}
  \label{fig:n101-akr}
\end{figure}

The left panel of Fig.~\ref{fig:n101-akr} displays the probability
distribution for the case $\sqrt{N}=41$ after 80 steps, which
corresponds to the first maximum of the probability at the marked vertex. The
running time is the instantaneous mixing time using as reference
distribution the one with maximum probability at the marked vertex and
taking a small value of $\epsilon$. This fact cannot be used to
estimate the running time, because there are no clues about how to
find that specific probability distribution beforehand.  On the other
hand, there is a method to find \textit{a priori} the limiting distribution.  Hence, an
interesting question is whether the limiting distribution can be used
to estimate the running time. The right panel of Fig.~\ref{fig:n101-akr}
displays the stationary distribution. Note the remarkable similarity between the distribution
with maximum probability and the stationary distribution.

\begin{figure}[h]
  \centering
  \includegraphics[width=0.45\textwidth]{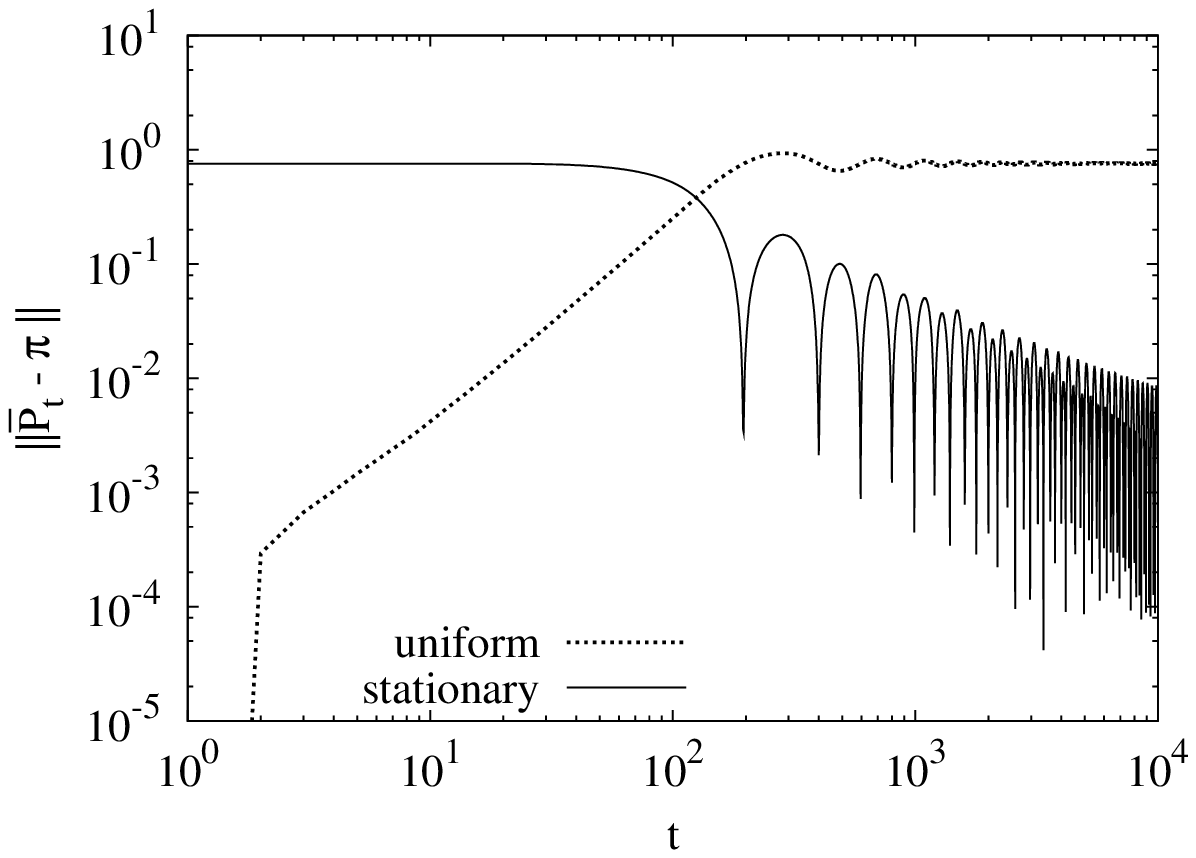}~
  \includegraphics[width=0.45\textwidth]{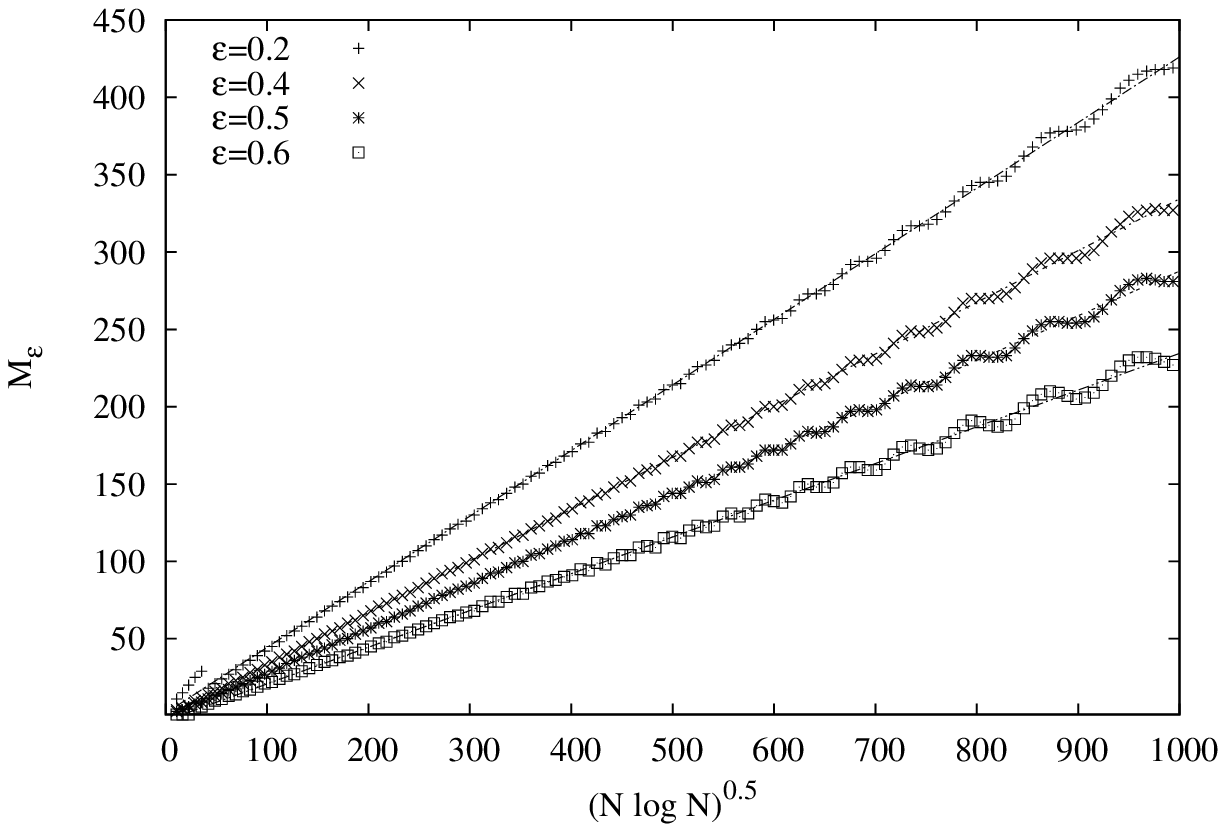}
  \caption{Left panel: total variation distance to both the uniform
    and the stationary distributions of the quantum walk in two-dimensional grid
    with modified coin used to search for a marked node.  Right panel:
    mixing time to the stationary distribution as a function of the
    input size.}
  \label{fig:akr}
\end{figure}

The first thing we need to check is how the mixing time of the modified walk scales with
$N$.  The right panel of Fig.~\ref{fig:akr} strongly suggests that replacing $C$ by $C'$ defined in Eq.~(\ref{Cprime}) does not alter the scaling of the average mixing time, since for $N\gg 1$ it scales as in eq.~(\ref{eq:Mepsilon2}).
There is a small oscillation in the mixing time function for large $N$, related to the way the modified walk
approaches the limiting distribution. In the the left panel of Fig.~\ref{fig:akr}, we have the total variation distance to both the uniform and the stationary distribution. For long times, the variation
distance approaches the stationary distribution with a strong
oscillation and several local minima, corresponding to the steps when
the instantaneous probability distribution has a maximum at the
searched node.

This behavior is remarkably different from the one observed in the
last section for the Grover walk on the two-dimensional grid (Fig.~\ref{fig:preakr}).
In that case the mixing time increases without oscillations and the total variation distance decays almost as a power law.

\section{Discussion}
\label{sec:conclusion}

In this paper, the mixing properties of a discrete-time quantum walk
on the finite two-dimensional grid with torus-like boundary conditions
has been considered in detail. This particular topology is quite
interesting due to its algorithmic applications.  The relation between
mixing time and the time complexity of the corresponding abstract
search algorithm has also been investigated.

The eigenvector problem of the evolution operator for the coherent
quantum walk has been solved and the corresponding stationary
distribution has been found analytically for the particular case of
odd lattices and a localized initial condition with a uniform superposition of coin states. The stationary
distribution is not uniform and has a maximum at the origin---for
which we have also provided a simplified expression. According to our
numerical simulations, the mixing time $M_\eps$ on the two-dimensional
grid with $N$ vertices increases as $O\left( \frac{\sqrt{N\log N}}{\eps^c} \right)$, i.e., almost quadratically faster than the mixing time of the classical random walk, which is $\Theta(N)$. The value of $c$ is approximately 1.  The faster mixing rates on the quantum
case is an advantage of using quantum-walk based algorithms over the
classical equivalents.

We have also considered a quantum walk in which the coin is modified
according to the prescription of the abstract search algorithm. We
have numerically calculated its limiting distribution and its mixing
time. Our numerical simulations show that the mixing time $M_\eps$ for
this particular walk also increases as $O\left( \frac{\sqrt{N\log N}}{\eps^c}
\right)$, where $c$ is approximately 1. This mixing time of the modified quantum walk
corresponds to the time complexity of marking a searched vertex on the
torus using the abstract search algorithm~\cite{Tulsi2008}. This
result establishes a relation between the mixing time of the modified walk and the running time of the corresponding
abstract search algorithm---which previously has been associated only
to the hitting time. This relation may be useful for the complexity
analysis of quantum-walk based algorithms.

In order to obtain the exact running time of the quantum algorithm,
one should also consider the constants that have been absorbed by the
asymptotic notation used to represent the mixing time. As a
future work, it would be interesting to investigate the mixing time of
search algorithms on other graphs and compare the corresponding
constants.

\begin{acknowledgments}
  The authors acknowledge financial support from CNPq (Brazil) and
  PEDECIBA (Uruguay).
\end{acknowledgments}

%\bibliographystyle{h-physrev} % PHYS REV adapted for arXiv preprints
%\bibliography{grid2d}

\end{document}